\RequirePackage{etoolbox}
\csdef{input@path}%
{%
 {sty/},
 {img/},
}
\documentclass{elsevierbook}

\usepackage{bm}
\usepackage{amsmath}
\usepackage{hyperref}

\usepackage{natbib}

\begin{document}

\def\aj{AJ}%
\def\araa{ARA\&A}%
\def\apj{ApJ}%
\def\apjl{ApJL}%
\def\apjs{ApJS}%
\def\ao{Appl.~Opt.}%
\def\apss{Ap\&SS}%
\def\aap{A\&A}%
\def\aapr{A\&A~Rev.}%
\def\aaps{A\&AS}%
\def\azh{AZh}%
\def\baas{BAAS}%
\def\jrasc{JRASC}%
\def\memras{MmRAS}%
\def\mnras{MNRAS}%
\def\pra{Phys.~Rev.~A}%
\def\prb{Phys.~Rev.~B}%
\def\prc{Phys.~Rev.~C}%
\def\prd{Phys.~Rev.~D}%
\def\pre{Phys.~Rev.~E}%
\def\prl{Phys.~Rev.~Lett.}%
\def\pasp{PASP}%
\def\pasj{PASJ}%
\def\qjras{QJRAS}%
\def\skytel{S\&T}%
\def\solphys{Sol.~Phys.}%
\def\sovast{Soviet~Ast.}%
\def\ssr{Space~Sci.~Rev.}%
\def\zap{ZAp}%
\def\nat{Nature}%
\def\iaucirc{IAU~Circ.}%
\def\aplett{Astrophys.~Lett.}%
\def\apspr{Astrophys.~Space~Phys.~Res.}%
\def\bain{Bull.~Astron.~Inst.~Netherlands}%
\def\fcp{Fund.~Cosmic~Phys.}%
\def\gca{Geochim.~Cosmochim.~Acta}%
\def\grl{Geophys.~Res.~Lett.}%
\def\jcp{J.~Chem.~Phys.}%
\def\jgr{J.~Geophys.~Res.}%
\def\jqsrt{J.~Quant.~Spec.~Radiat.~Transf.}%
\def\memsai{Mem.~Soc.~Astron.~Italiana}%
\def\nphysa{Nucl.~Phys.~A}%
\def\physrep{Phys.~Rep.}%
\def\physscr{Phys.~Scr}%
\def\planss{Planet.~Space~Sci.}%
\def\procspie{Proc.~SPIE}%

\Frontmatter
  \include{tableofcontents}

\Mainmatter
  \begin{frontmatter}

\setcounter{chapter}{9}
\chapter{Coronal Heating}\label{chap10}


\begin{aug}
\author[addressrefs={ad1,ad2}]%
  {\fnm{I\~nigo}   \snm{Arregui}}%
\author[addressrefs={ad3}]%
 {\fnm{Tom} \snm{Van Doorsselaere}}%

\address[id=ad1]%
  {Instituto de Astrof\'{\i}sica de Canarias, V\'{\i}a L\'actea s/n, E-38205 La Laguna, Tenerife, Spain}%
 
\address[id=ad2]%
  {Departamento de Astrof\'{\i}sica, Universidad de La Laguna, E-38206 La Laguna, Tenerife, Spain}%
  
\address[id=ad3]%
  {Centre for mathematical Plasma Astrophysics, Department of Mathematics, KU Leuven, B-3001 Leuven, Belgium}%

\end{aug}

\begin{abstract}
Coronal heating refers to the physical processes that shape and structure the corona of the Sun and are responsible for its multi-million Kelvin temperatures. These processes are revealed in a number of different observational manifestations and have been studied on theoretical grounds in great detail over the last eight decades. The aim of this Chapter is to give an account of some of those manifestations and to discuss relevant physics that we believe is responsible for them. Coronal heating is closely connected to other magnetohydrodynamic (MHD) processes occurring in the solar plasma and described in this book such as waves, shocks, instabilities, and magnetic reconnection.
\end{abstract}
\begin{keywords}
\kwd{Sun: corona}
\kwd{Sun: magnetic fields}
\kwd{Magnetohydrodynamics (MHD)}
\end{keywords}

\end{frontmatter}

\section{Introduction}

The outermost layer of the solar atmosphere, the corona, is made-up of plasma with a very high electrical conductivity, and is permeated by a highly structured magnetic field displaying a wide range of spatial scales. Plasma and magnetic field interactions are responsible for phenomena like the heating of the solar corona or the acceleration of the solar wind. They are the ultimate source for Sun-Earth interactions. Gathering a satisfactory description of these phenomena still represents a challenge to our understanding of the Sun, and prevent us from mitigating their effects on Earth.

The so-called coronal heating problem originated about eight decades ago, when Edl\'en and Grotrian identified Fe {\sc IX} and Ca {\sc XIV} spectral lines in the corona, indicating the presence of fully ionised plasma at 1 million degree Kelvin \citep{edlen43}. Since then, solar physics research has sought to identify the physical processes that might balance the three primary coronal energy loss mechanisms; thermal conduction, radiation, and solar-wind outflow, first quantified by \cite{withbroe77}. 

In spite of decades of research, the coronal heating problem remains unsolved \citep[see][for reviews than span across different decades]{kuperus69,withbroe77,kuperus81,aschwanden05b,klimchuk06,parnell12,klimchuk2015,vandoorsselaere20}. Researchers in the field agree upon the fact that the ultimate reason for the multi-million degree temperatures in the outer solar atmosphere is of magnetic nature and that the energy source lies in the convective motions at the photosphere. The exact physical processes and their relative contribution to the heating of the plasma remain largely unknown or unquantified.

Two main types of processes are believed to contribute to the heating: the direct dissipation of magnetic fields by mechanisms such as magnetic reconnection \citep{sturrock81}, current cascades \citep{parker63}, viscous turbulence \citep{vanballegooijen86}, or magnetic field braiding \citep{peter04}; and the dissipation of magnetic wave energy stored, transported, and transferred to short spatial scales by magnetohydrodynamic (MHD) waves \citep{alfven47,ionson78,heyvaerts83}.
There is ample evidence for the occurrence of both magnetic reconnection \citep{zweibel09} and magnetic wave dynamics \citep{nakariakov05}. They have both been studied extensively, and significant theoretical advancements on their description were made, so a discussion in terms of mutually exclusive mechanisms is beside the point. It is however important to assess the relevance of each mechanism in heating the plasma of the solar atmosphere.

The aim of this Chapter is to give an account of some observational manifestations of coronal heating and a discussion on the physical processes that are believed to be most relevant in causing them.

\section{Coronal emission}

The corona of the Sun is the outermost part of the solar atmosphere. It extends just above the chromosphere towards the interplanetary medium.  Light emitted from the white-light corona shows up as a faint and extended halo, with an intensity near the edge of the solar disc that is about a million times fainter than the white-light coming from the disc. The light scattered by the Earth's atmosphere is several times brighter that the corona.  This makes impossible to observe the white-light corona from Earth unless the disc of the Sun is completely covered. The corona is visible to the naked eye at times of total solar eclipses, when the solar disc is obscured by the Moon (see Figure~\ref{fig:coronalemission}). 

The shape of the corona is generally irregular and varies in time, depending on the stage of the solar magnetic cycle. 
Near or at sunspot maximum, the corona shows an irregular lobe-form extending around the lunar disc with no preferred solar latitude, although the intensity is reduced near the poles. At sunspot minimum, the corona appears symmetrically elongated about the equator, because active regions are located at low latitudes. It displays long streamers, extending from opposite sides of the Sun and approximately aligned with the solar equator, as well as polar plumes.

The origin of coronal white-light emission is the scattering of light emitted from the photosphere. The scattering is due to both free electrons and dust grains.  In terms of the different processes that cause coronal emission, we can distinguish the following components:

\begin{figure}
 \title{Coronal emission and structure}
  \includegraphics[width=6.05cm]{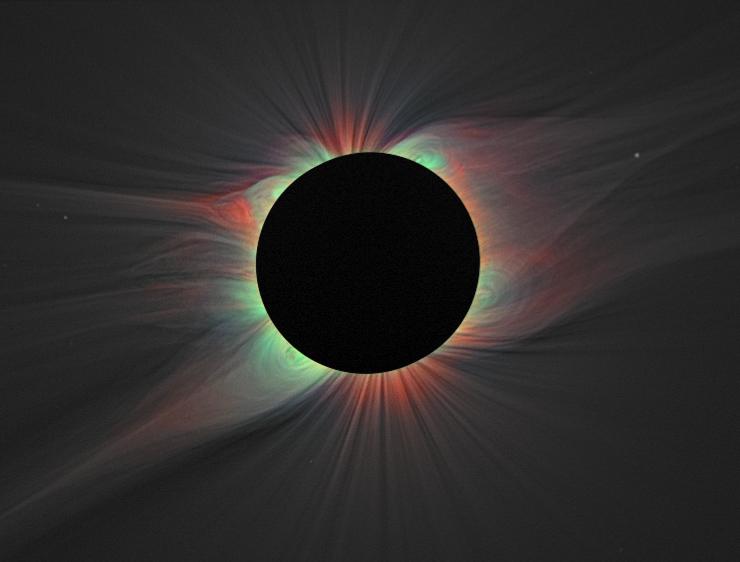} 
  \includegraphics[width=4.6cm]{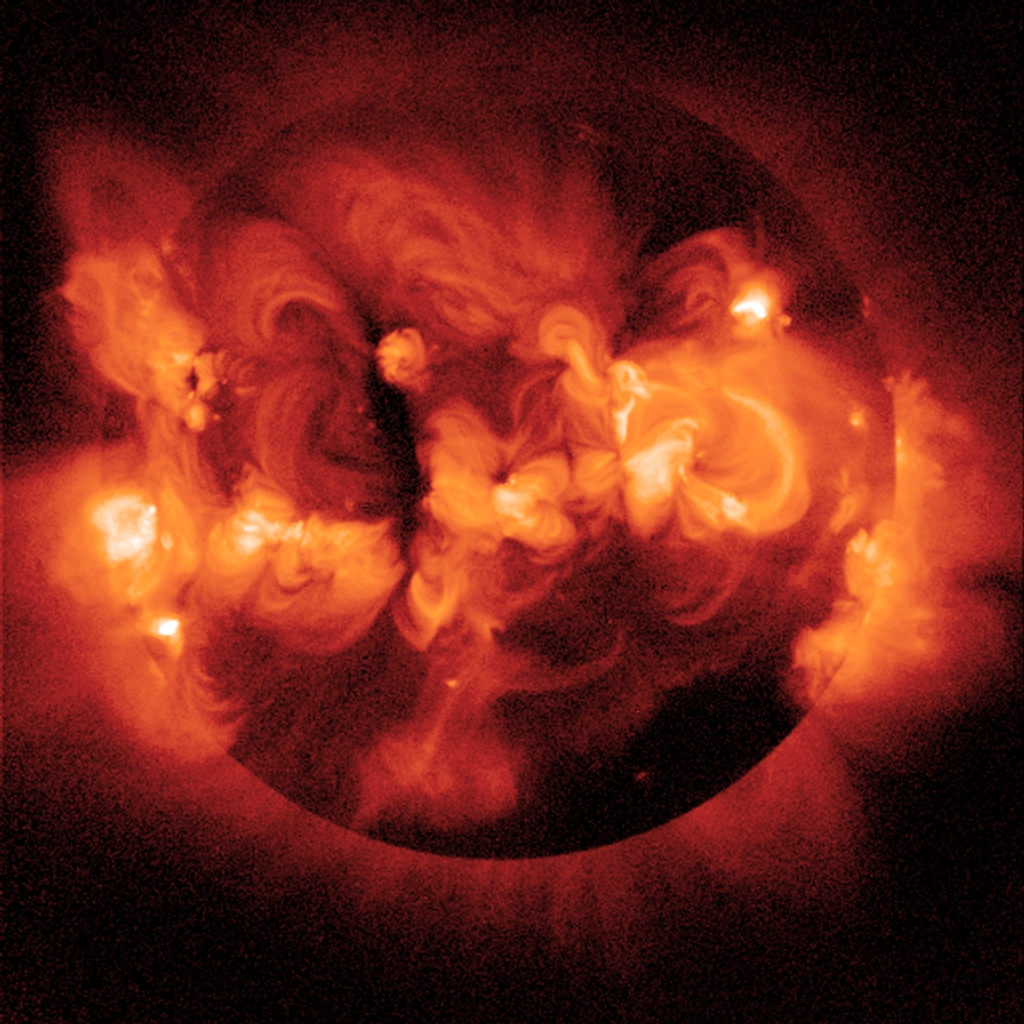}
 \caption{On the left, image of the solar corona taken during a total solar eclipse. Red indicates the Fe XI 789.2 nm line, blue the Fe XIII 1074.7 nm line, and green shows the Fe XIV 530.3 nm line. Credit: Habbal/NASA. On the right, X-ray image taken by Yohkoh satellite on 1February 1992 01:44 UT, showing the variable brightness distribution of the corona due to its magnetic structuring, which gives rise to different manifestations of solar activity. Solar X-ray image courtesy of the Yohkoh Legacy data Archive
(http://solar.physics.montana.edu/ylegacy) at Montana State University.} 
 \label{fig:coronalemission}%
 \end{figure}

\begin{itemize}
\item The K-corona originates from Thomson scattering of photospheric light by free electrons. It displays a featureless continuous emission spectrum like that of the photosphere, but without Fraunhofer lines and is strongly polarised. The absence of Fraunhofer lines is due to the high temperatures and low densities, making electrons travel at very high speeds in the corona. At a temperature of 2 MK, the electrons move with an average speed of 10~000~ km~s$^{-1}$. As a consequence, the Fraunhofer lines of the photospheric spectrum are extremely wide and no longer perceptible as such.

\item The F-corona arises due to scattering of the Fraunhofer spectrum by the dust particles in the interplanetary atmosphere. It is more an extension of the zodiacal light, rather than a genuine solar phenomenon.  This emission shows no polarisation. Because the dust grains move at a much lower speed than the electrons causing the K-corona, the Fraunhofer lines remain visible in the F-coronal emission.

\item The E-corona arises due to the high temperatures, of the order of a few MK and the low densities around $\sim$10$^{10}$ cm$^{-3}$ in the corona and is the only component produced by the coronal gas itself. Under such extreme conditions,  ions produce emission lines. The total integrated brightness of the E-corona is small in comparison to the K- and F-coronae. However, the strong emission relative to the background scattered light, concentrated around specific wavelength, makes possible to observe coronal features through narrow-band filters centred in these emission lines. 

\end{itemize}

The existence of a hot corona can already be deduced from the apparently simple observational manifestations described above. Because there is essentially no dependence on temperature, the intensity is proportional to the line-of-sight integral of the density. The surface brightness of the K-corona depends on the electron number density. Along a given line-of-sigh through the corona, the surface brightness relative to the photospheric brightness can be calculated as the number of electrons times the electron cross-section. Following \cite{phillips95}, consider a typical coronal streamer of about 1000 km extent. This leads to a density of about 10$^{14}$ electrons per cubic metre. The electron density varies with distance from the Sun, but decreases relatively slowly with distance. This is again an indication of high temperature. Using pressure scale arguments and estimating the characteristic length-scale over which density decreases in data for surface brightness variation \citep{allen73} temperatures as high as 2 000 000 K are obtained \cite{blackwell67}.  

The coronal spectrum was known already in the 40s of the past century, but its properties were not well understood. The main puzzles were the absence of Fraunhofer lines in the bright continuum observed close to the Sun's limb (inner corona), the presence of unidentified bright emission lines, the absence of emission lines of H and metal,  the width of coronal emission lines, and the extended size of the corona. It turns out that they are all indications of high kinetic temperature plasma.

A strong indication of the hot corona is the presence of so-called ``forbidden" lines in coronal emission and their identification (by Edl\'en and others) with highly ionized ions, such as the coronal  ``green" Fe XIV and ``red"  FeX lines at 530.3 and 637.5 nm, respectively. \cite{edlen43} identified coronal bright emission lines as being due to elements, such as A, Ca, Fe, Ni, in very high states of ionisation. Very high temperatures and low densities were required to explain the presence of those lines with such ionisation degrees in the coronal spectrum. The intensities of these lines depend on temperature, which controls how much of a particular ion is present relative to all the other ions of an element and also causes the excitation from the ground state to the upper state from which de-excitation leading to the emission is produced. Plasma density is relevant because without low densities electron collisions would produce de-excitation/excitation to another states, preventing the forbidden line to be observed. 

Observations of the corona in visible wave-lengths are limited by the occurrence of eclipses or the use of coronagraphs. In addition, they do not permit us to observe coronal features on the disc because the corona is optically thin at visible wavelengths and optical radiation is not absorbed.  The situation is different for radiation in non-visible wavelengths, such as X-rays, extreme ultraviolet (EUV) or radio wavelengths, which are emitted by the corona but not at all by the photosphere. The fact that the primary emission from the corona is in EUV and X-ray bands is another clear indicator of the multi-million K coronal temperatures. 

\section{Coronal structure}

Solar (and stellar) EUV and X-ray emission requires the existence of a magnetic-field-generating dynamo process, which provides the means to connect the convection zone with the atmosphere. Magnetic fields generated within the convective layer feed mechanical energy into the atmosphere and produce plasma heating and structuring.

As such, EUV and soft X-ray images of the solar corona show that the brightness distribution is far from uniform (see Figure~\ref{fig:coronalemission}). In these wavelengths, coronal features become observable against the solar disc.  A highly structured corona becomes apparent, which displays three major types of features: (a) active regions and coronal bright points, which are characterised by strong magnetic fields and closed field configurations; (b) coronal holes, characterised by relatively weaker field strengths and open field configurations; and (c) quiet Sun coronal regions, in which the field is closed, but over larger spatial scales. Each of these features is, in turn, characterised by a different level of X-ray and EUV intensity, clearly indicating the important role of the magnetic field in the energy balance and heating of the corona. The magnetic field structure conditions the geometry of mass and energy flows, guides magnetic waves, and stores magnetic energy that can be released into the surrounding plasma.

\begin{figure}
 \title{Coronal emission and the solar activity cycle}
 \includegraphics[width=8cm,angle=-90]{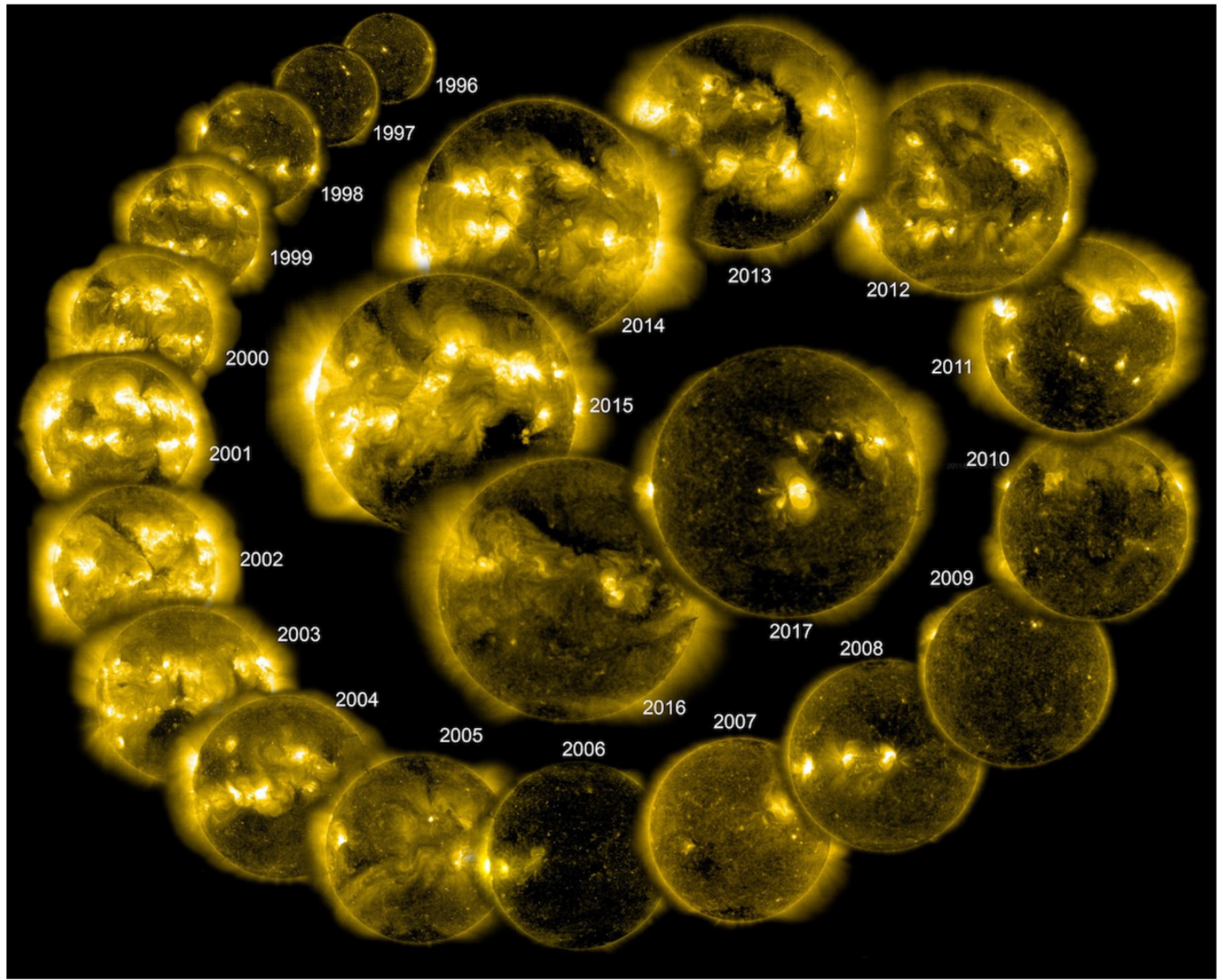}
 \caption{A collage of images displays coronal emission in extreme ultraviolet light at different stages of the solar cycle (every spring from 1996 to 2016) taken with the Extreme ultraviolet  Imaging Telescope (EIT) onboard the Solar and Heliospheric Observatory (SOHO). From \cite{temmer21}.}
 \label{fig:cycle}%
 \end{figure}

The structure of the corona in EUV and soft X-ray images is determined by the structure of the magnetic fields that permeate the plasma which, in turn, depends on the cyclic variation of the solar activity (see Figure~\ref{fig:cycle}). EUV images near solar maximum display a large fraction of the surface covered with active regions located at two well-defined latitude bands above and below the equator. At solar minimum, EUV emission is in general fainter and active regions are scarce. They show up at higher latitudes. At these times, X-ray emission is dominated by smaller magnetic bright points. Open-field regions lead to the so-called coronal holes, which show up at the polar caps during most of the cycle. They have their largest extent during solar minimum and can appear crossing the equator in between the minimum and maximum of the solar cycle. Any plausible coronal heating mechanism, should provide an explanation to these general features and their variation with the solar-activity-cycle.

\section{Coronal density and temperature distribution}\label{sect:rhotemp}

Gaining knowledge about the physical processes behind the heating of the corona requires the assessment of its physical conditions; most significantly its density, temperature, and their spatial and temporal variation. Early efforts at this respect are reviewed in \cite{allen54}.

The density distribution of the corona can be obtained from the scattering of light by electrons. Initial studies were able to deduce this density distribution in the form of a sum of powers of radial distance, from the projected distribution of light seen from the Earth, and assuming a given electron distribution in the corona,  see, e.g., \cite{vandehulst50}. The obtained values were mean densities, averaged from observations taken during several eclipses and over volumes that were large enough to include several coronal features. 

Early coronal temperature estimates were based on the width of coronal emission lines. The absence of hydrogen and metals emission lines indicated that the electron temperature must be above 700 000 K \citep{goldbergmenzel48}. The observations of \cite{grotrian31} led to temperature estimates of about a million degrees. Values of several million K were reported by  e.g., \cite{lyot37} and \cite{waldmeier44}. Early methods for estimating coronal temperatures involved the analysis of Doppler broadening of emission lines; line ratios from multiple states of ionisation of ions; the estimate of the density gradient resulting from intensity measurements; the analysis of radio emission at 10-cm wavelength in quiet-corona regions; or the degree of ionisation from ionisation cross-sections and recombination coefficients for the ions concerned \citep[see e.g.,][for example]{brosius94,noci03}. More recent techniques, used either as an alternative or in combination with those, infer the coronal temperature from observed properties of MHD slow-wave modes, using seismology inversion techniques \citep{marsh09a,marsh09b}.

The density and temperature of the coronal plasma vary in space and time. Based on the spatial distribution of the temperature, the corona is classically divided into three regions: the lower corona, the region closest to the Sun's surface, has a temperature of around 1 MK and a relatively high density of about 10$^8$ particles per cubic centimeter. The region just above, the intermediate corona, has a  temperature of around 1-2 MK. The density in this region is lower than in the lower corona, around 10$^6$ particles per cubic centimeter. The upper corona is the region farthest from the Sun's surface and has a temperature of several MK. The density in this region is extremely low, around 10$^4$ particles per cubic centimeter. 

 \begin{table}[tbp]
 \title{Coronal energy losses [W m$^{-2}$] from \cite{withbroe81}.}
 \centering
 \begin{tabular}{lccc}
 {\sc Energy loss mechanism} & {\sc Quiet Sun} & {\sc Active Region} & {\sc Coronal Hole}\\
 \hline
 Thermal conduction flux & 2 $\times$ 10$^2$ & 10$^2$ -- 10$^4$ &60\\
 Radiative flux & 10$^2$ & 5 $\times$ 10$^3$ &10\\
 Solar wind flux & <  50 & < 10$^2$ &7 $\times$ 10$^2$\\
 {\sc Total} &{\sc 3 $\times$ 10$^2$} & {\sc 10$^4$} &{\sc 8 $\times$ 10$^2$}\\
 \end{tabular}
 \caption{}
 \label{table:losses}%
 \end{table}

Figure~1.2 in \cite{golub09} shows the classical schematic representation of the temperature and density variations in the solar atmosphere as a function of height. This is the result of semi-empirical models, based on the computation of solutions to radiation transfer equations for hydrogen, carbon, and other constituents. The solutions are computed 
on spherically symmetric and homogeneous plane-parallel atmosphere models, with physical properties varying with height. They consider optically thick plasmas in non-local thermodynamic equilibrium conditions \citep[see][for details]{vernazza81}.   Recent imaging observations of the corona in EUV emission lines demonstrate that the adopted plane-parallel view is extremely simplistic, because of the highly structured nature of the corona they display. It is still of interest to know what the key features are in temperature and density dependence in those simplified atmospheric models. They form the basis for many current numerical computations for the dynamics of the coronal plasma \citep[e.g.][]{howson22}.  \cite{fontenla93} computed energy balance hydrostatic models of the atmospheric parameters, which are a continuation of the semi-empirical models by \cite{vernazza81}. Nowadays, a more accurate definition is one that considers the solar corona as any portion of the solar atmosphere with a temperature above 10$^{5}$ K \citep{golub09}.

These physical conditions completely determine the energy balance on any given volume of the solar corona. Energy balance estimates for different components of the corona are shown on Table~\ref{table:losses}, with values taken from Table 2.1 in \cite{withbroe81} (see also Table 1.1 in \citealt{golub09}). They consider the amount of energy loss by three mechanisms: thermal conduction, radiation, and plasma outflow.  The estimates strongly depend on the particular region of the corona being considered and also on the strength and configuration of the magnetic fields. In coronal holes, the mean temperature and density are lower in comparison, hence conductive and radiative losses are less than in the active region corona. The open magnetic field lines enable large energy losses by plasma outflows, making the total losses higher than in quiet Sun regions. On the other hand, active region energy balance is mainly conditioned by energy lost by thermal conduction and radiation. Coronal energy losses in active regions exceed those in the quiet Sun and coronal hole regions by one order of magnitude, approximately.


\begin{figure}
 \title{The structure of coronal active regions}
 \includegraphics[width =10cm]{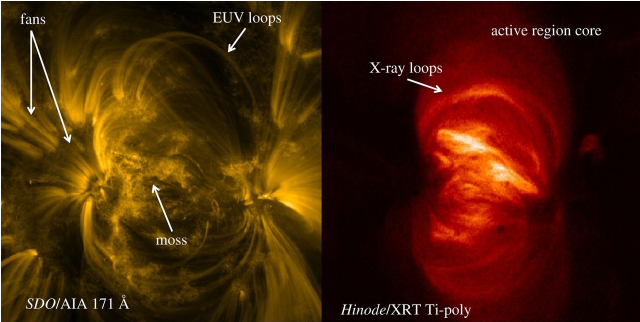}
 \caption{EUV image in the 171 \AA\ channel (1 MK) from SDO/AIA (left) and X-ray counterpart  from Hinode/XRT (T > 2 MK) showing the different components that conform a typical active region in the corona. From  \cite{schmelz15}.} 
 \label{fig:ARstructure}%
 \end{figure}

\section{Coronal active region loops}

The temperature and density distribution in the corona are influenced by the magnetic field. The magnetic-field lines in the corona trap plasma and create a highly structured corona, which is far from being homogeneous at different spatial scales. This magnetically dominated corona consist of a variety of structures, such as active regions, coronal loops, X-ray bright points, coronal holes, etc. Coronal loops are the basic building blocks of active regions, areas of the solar corona where the magnetic field is concentrated. They show up as regions with enhanced extreme ultraviolet and X-ray emissions and their temperatures are higher than the surrounding quiet Sun corona. They have a particular morphology and are built by components with different geometrical and physical properties (see Figure \ref{fig:ARstructure}). 

The central area of an active region is the core, which appears in X-ray wavelengths as a diffuse area where individual loops are difficult to isolate from observations. It is formed by short loops with temperatures in the range of 3 to 5 MK. In the hotter emission lines the loops appear more fuzzy. At the foot-points of these loops, a bright reticulated pattern is observed in EUV emission, the so-called moss. This structure appears to closely follow the photospheric magnetic field distribution. Surrounding the core, we find 1 MK more outlined and longer loops. These are seen to evolve in shorter time-scales in comparison to the core.  The periphery of the active region is filled with long and cool fan loops. The background is filled with diffuse emission, which is rather interesting because it is clearly heated, but not connected to strong magnetic field structures. 

\begin{figure}
 \title{Equilibrium of active region loops}
 \includegraphics[width =7cm,angle=-90]{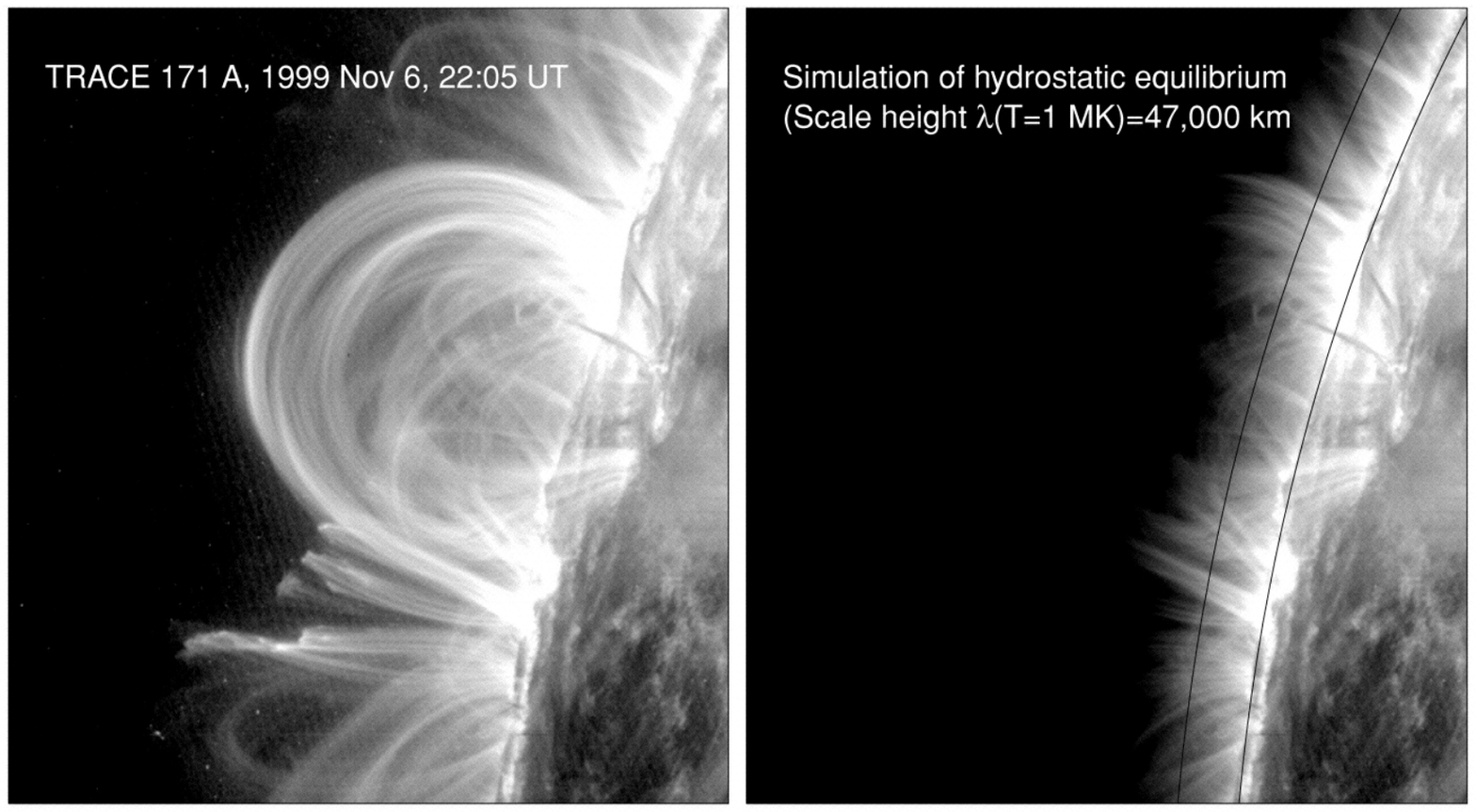}
 \caption{Transition Region and Coronal Explorer (TRACE) observation of an active region with several loops (left) and a simulation of the observed intensity scaled to the hydrostatic thermal scale height of T = 1 MK (right), with a pressure scale height of about 46 Mm. From \cite{aschwanden01loops}. Reproduced by permission of the American Astronomical Society (AAS).} 
 \label{fig:scaleheight}%
 \end{figure}

Physical properties discussed in Section~\ref{sect:rhotemp} refer to global conditions, average values that include several coronal features. The analysis of modern high-resolution imaging and spectroscopic observations enables us to differentiate the conditions in active regions and their components and offers means to diagnose the properties of plasma heating processes. Current estimates of the temperature and density in the corona are obtained by means of spectroscopic diagnostic techniques. They involve the forward modelling of spectral lines properties, such as their intensity, line-width, and profile. By measuring the intensities of spectral lines over a range of temperatures, the differential emission measure can be determined as a function of the temperature (see Section~\ref{sect:emissionmeasure} below), which informs about the amount of plasma at a given temperature. Some ions have low-lying meta-stable levels whose populations are sensitive to the density in particular density ranges. Taking ratios of the intensities of those selected spectral lines can be used to determine plasma densities, since the ratio of the population of the ground and the meta-stable levels will be density sensitive. Line ratios can also be used to obtain information on the temperature, considering two spectral lines with differences in excitation energies from a common ground level. The broadening of spectral line profiles gives information on Doppler shift introduced by, e.g., mass flows, waves, or unresolved turbulent velocities. 


Observational analyses of this kind, made with the first imaging observations of coronal loops in EUV wavelengths with TRACE,  made apparent a few observational issues. For example, coronal loops are nearly isothermal and have almost constant temperature distributions along their coronal segments, as already noted by \cite{withbroe77}. This is manifested as relatively flat 195 to 171 \AA\ filter ratios along much of their lengths. The pressure scale-height at 1 MK is about 46 Mm, hence long loops should be gravitationally stratified. Imaging observations, on the contrary, display bright emissions many scale heights above the solar surface (see Figure~\ref{fig:scaleheight}). The observed flux is proportional to the emission measure, which has the half pressure scale height of $\sim 23$ Mm.  Hence, active region loops seem to be over-dense having enhanced intensities in comparison to the properties predicted by simple hydrostatic scaling laws governing the relationship between pressure, temperature, and loop lengths \citep{rosner78,serio81,aschwanden01loops,schmelz15}.  The observationally determined super-hydrostatic scale-heights are incompatible with steady uniform heating models \citep{winebarger03}. Hydrostatic loop models can only explain well the shortest loops \citep{aschwanden01loops,winebarger03}. On the other hand, models in which heating is localised at the footpoints lead to flatter temperature profiles and larger apex densities, in better accordance with observations \citep{winebarger03}.  For instance, \cite{warren02} suggested that impulsive heating can produce loops that are overdense relative to the predictions of uniformly heated static loop models.

Another striking observed feature is the persistence of loop's bright emission in timescales much longer than the characteristic radiative cooling times in the corona \citep{lenz99a,lenz99b}. \cite{winebarger03b} studied the temporal evolution of five active region loops observed with TRACE. The loops appear first in the 195 \AA\ filter and then in the 171 \AA\ filter. The progression in the appearance of loops from the hotter filters to the cooler filters is consistent with the cooling loops model proposed by \cite{warren02}. 

\begin{figure}
 \title{Hydrodynamic evolution of coronal loops}
 \includegraphics[width =8cm,angle=0]{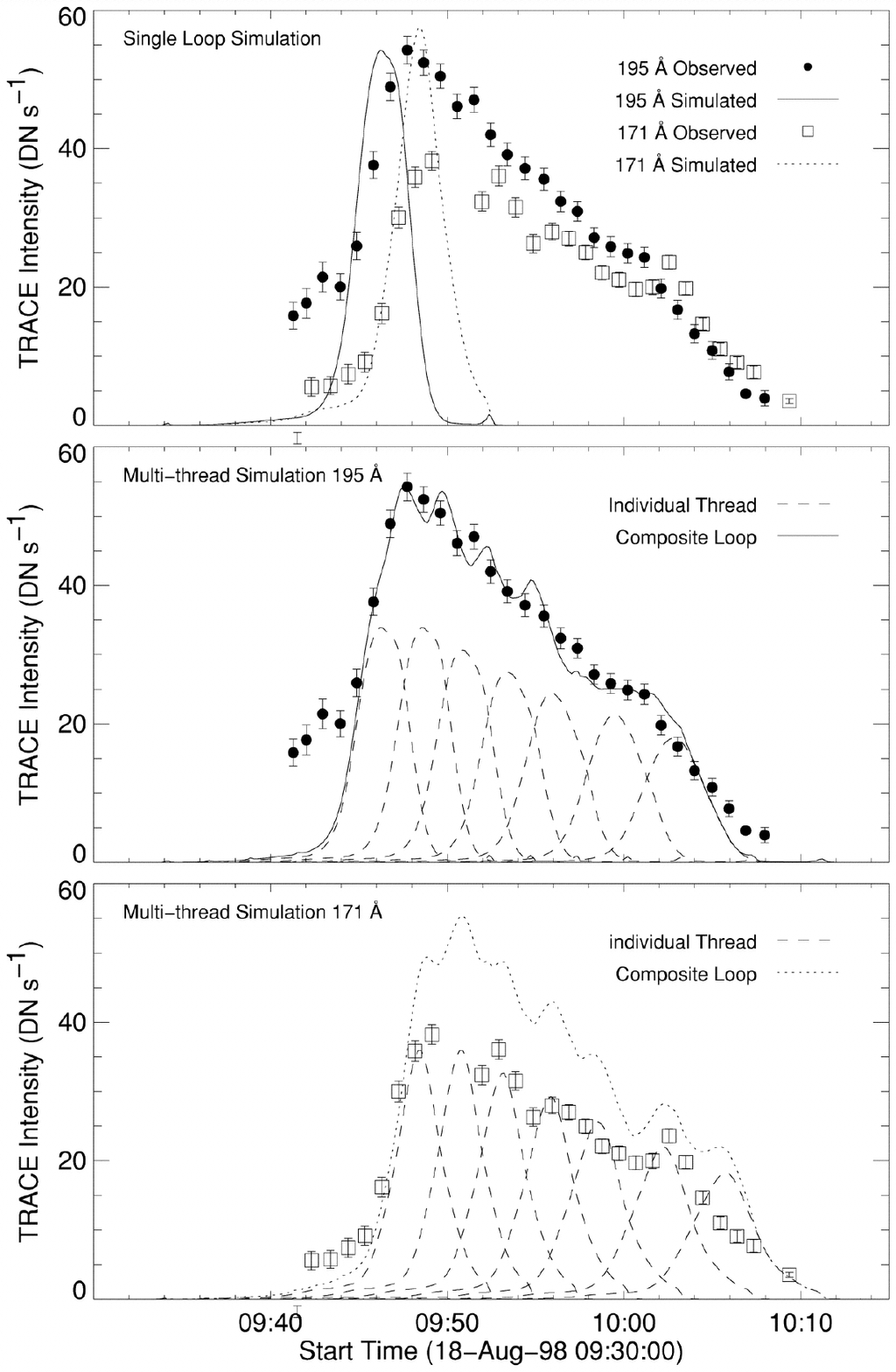}
\caption{Top panel: a comparison between simulated and observed light curves with data from the Transition Region and Coronal Explorer (TRACE) in an active region loop observed on 18 August 1998.  In the single loop simulation, the delay between the 195 and 171 \AA\ intensities in the simulated light curve matches the observations, but the loop cools too quickly. In the Middle and Bottom panels, simulated light curves arise from models with a series of threads that are heated sequentially and these models are able to reproduce better some observed characteristics of the observed light curves. From \cite{warren03}. }
 \label{fig:hydroevollopps}%
 \end{figure}

\cite{warren03} performed detailed hydrodynamic modeling of one of the TRACE loops analyzed by \cite{winebarger03b} and compared observed and simulated light curves for the evolution of coronal loops. Some properties of hydrodynamic simulations, such as the cooling and draining times, relate to features of the observed light curves, such as the delay between the appearance of the loop in the different filters. An example is displayed in Figure~\ref{fig:hydroevollopps}, which shows the comparison between observed light curves in 171 and 195~\AA\ with simulation results from monolithic and multi-thread models \citep{warren03}. A loop lifetime comparable with the cooling time is compatible with a monolithic (isothermal) loop structure. On the other hand, the observed relatively flat filter ratios and long lifetimes of coronal loops can only be obtained in simulations by considering an ensemble of independent, impulsively heated strands, even if they are ``seen'' as a single loop in imaging observations. In general, it is possible to match both the spatial and temporal evolution of the observed loops by assuming that they are actually a collection of threads that are heated sequentially. 

\cite{winebarger04} showed that the analysis of impulsively heated 1 MK active region loops, as they cool down and show up in different bandpasses, can be used to determine crucial characteristics, such as the magnitude, duration, and location of the energy release.  The evolution of the apex density and temperature in such loops depends only on the total energy deposited. This would mean that observations must be made early in the evolution of a loop to be able to determine the heating parameters and to discriminate between different heating scenarios.

\subsection{Intensity variations}

A widely used observational diagnostic tool to understand how coronal loops are heated is based on the analysis of the intensity variations of measurements at different EUV or X-ray wavelengths, which is then related to the thermal evolution of the plasma.  As first shown by \cite{winebarger05} \citep[see also][]{ugarteurra06,warren07,ugarteurra09,viall11,reale14},  some coronal loops appear first in X-ray emission (T >  2.5 MK) before showing up as bright structures in EUV imaging observations.  This would  imply a cooling process in which the peak emission at different wavelengths is reached at consecutive times as the loops' temperature goes  down. Improvements in spatial and temporal resolution, as well as the increase in the number of EUV channels, provided by the Solar Dynamics Observatory (SDO: \citealt{lemen12}), have enabled measuring the time-delay between those consecutive peak times, making possible the observational diagnostics of heating/cooling processes in coronal loops and giving information about the frequency of heat deposition.  

The concept of frequency of heat deposition is based on the so-called Parker nanoflare-hypothesis \citep{parker88}, which postulates that plasma heating in the corona is the result of microscopic and unresolved magnetic reconnection events, uniformly distributed over the full coronal volume. Under this hypothesis, low-frequency heating refers to a scenario in which the repeat time of heating events is much longer that the plasma cooling time (impulsive heating regime). Conversely, high-frequency heating refers to the opposite scenario in which the repeat time is much shorter than the plasma cooling time, which would essentially be akin to a steady heating process.   In the intermediate case, plasma cooling and heating repeat times can be of the same order. Much of the observational analyses in recent years have focused on the inference of the impulsive/steady character of the underlying heating mechanism from the analysis of X-ray/EUV imaging and spectroscopic observations.  

A key idea is that when such a small-scale heating event (nanoflare) occurs, the heating phase contributes little to the observed emission and it is the cooling phase which dominates the observed light curves \citep{bradshaw11}. By means of a time-lag analysis, \cite{viall12} computed the cross-correlation between light curves in six successively cooler SDO/AIA EUV channels (131  \AA, 94 \AA, 335 \AA, 211 \AA, 193 \AA, and 171 \AA) at given locations of an active region. The resulting patterns show intensity peaks in the successively cooler pass-bands being sequentially reached, as the plasma cools through the sequence of EUV wavelengths. \cite{viall15} extended the analysis to the quiet Sun and later \cite{viall17} applied this type of analysis to 15 active regions catalogued by \cite{warren12}. Their results indicate overwhelmingly positive time lags (cooling plasmas) in all cases, with only a few isolated instances of negative time lags (heating plasma).  These  observations are interpreted as a manifestation of a impulsive heating scenario in which little emission is produced during the heating phase and observations are indicative of cooling of plasma that has been impulsively heated. Some areas show a time lag of zero which, rather than lack of variability, are interpreted as a manifestation of strong variability consistent with the response of the transition region to heating events \cite[see also][]{viall16}.

A number of subsequent studies have applied the same method to simulated synthetic images obtained from models. For instance, \cite{bradshaw16} find that some aspects of the observed light-curves can be reproduced by both high and intermediate-frequency nanoflare models. On the other hand,  \cite{lionello16} were unable to reproduce the time-lag characteristics from \cite{viall12} in their field-aligned hydrodynamic models. This means that either the time delays may not be representative of the real loop evolution, or that other heating scenarios beyond the impulsive heating and cooling must be considered to explain the observations.
Overall, impulsive heating is consistent with EUV loops \cite{ugarteurra06,hara08}, while steady heating would better explain the soft X-ray emission in active region cores \citep{warren10, winebarger11,tripathi11}. 

\subsection{Emission measure distribution}\label{sect:emissionmeasure}

\begin{figure}
 \title{The loci approach to assess the emission measure distribution}
 \includegraphics[width =6cm,angle=-90]{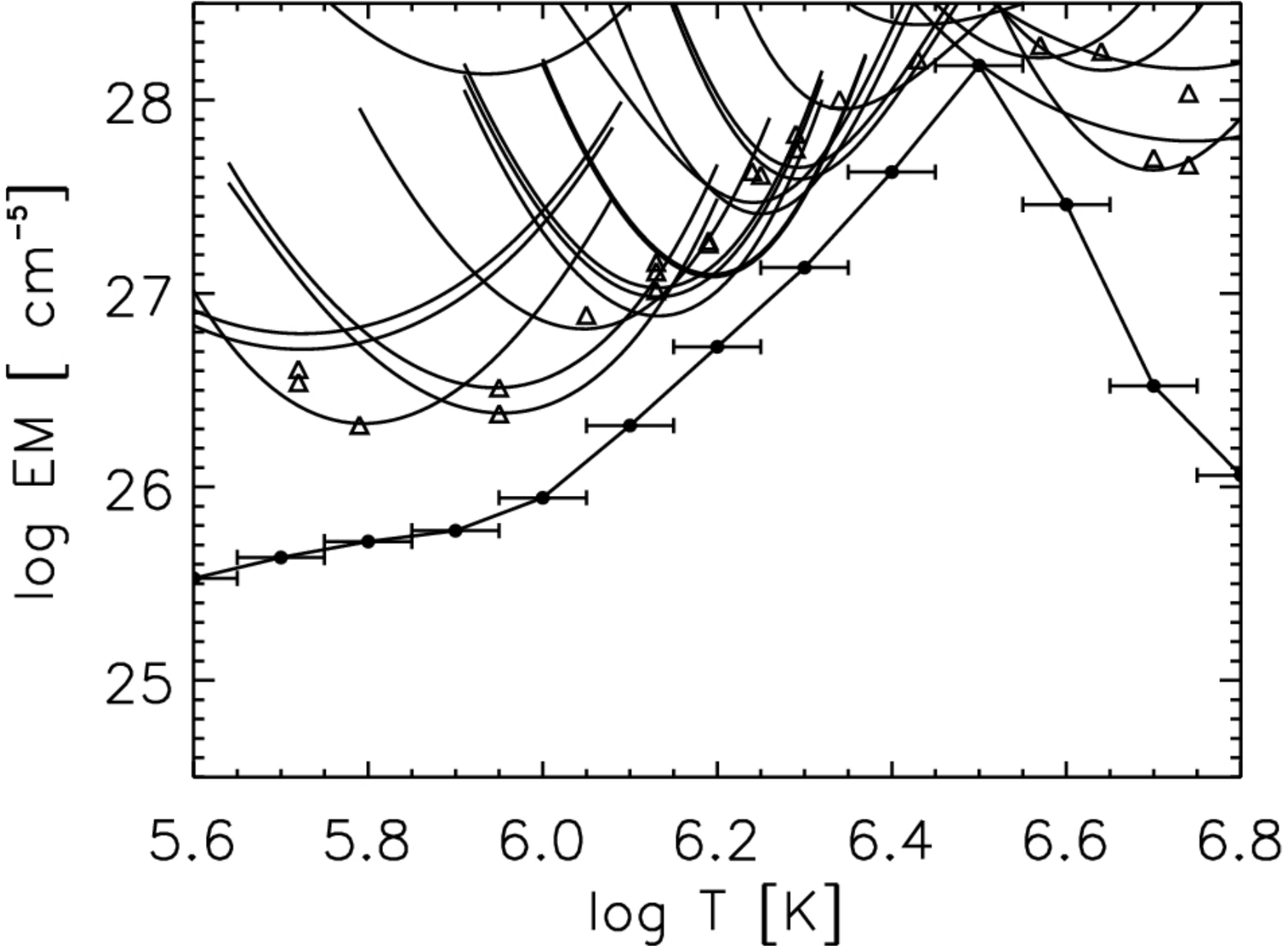}
\caption{Emission Measure (EM) loci curves for a selection of lines observed by Hinode/EIS in an active region core \citep{delzanna14}. The filled circles represent the values calculated from the Differential Emission Measure (DEM) with $\Delta \log T=0.1$ K. The triangles, plotted at the temperature $T_{\rm max}$, are the solution using the \cite{jordan71} approximation. From \cite{delzanna18}. }
 \label{fig:lociEM}%
 \end{figure}

Another method to diagnose the plasma physical conditions and the heating characteristics from observations is based on the analysis of the so-called emission measure (EM) distribution, $\mathrm{EM}(T)=\int {dh}\,{n}_{e}^{2}$, with $n_{\rm e}$ the electron density and the integration is taken alone the line of sight. This enables to determine the temperature distribution of the coronal plasma, with the expectation that this temperature structure can inform about how the corona is heated \citep{warren12}. The distributions are derived from the intensities at different spectral lines by computing the EM at the temperature that maximises the contribution function, the so-called peak formation temperature. For each spectral line, it is assumed that the contribution function is constant over a given temperature range around the peak formation temperature and zero otherwise. An example is shown in Figure~\ref{fig:lociEM}, which displays the resulting inverted U curves, the so-called EM loci plots. Each of these curves gives the amount of emission measure that would be needed to produce the observed intensity at each spectral line, if the plasma were isothermal at each temperature. The position of the minima corresponds to the peak formation temperature for each line, since this is the temperature at which the line emits more efficiently and, hence, demands least emission measure \citep[see][for a detailed description]{delzanna18}.

Emission measure distributions derived from observations can then be compared to those predicted by theoretical models. In these models, the cool portion of the EM(T) distribution, to the left of the peak, is usually described by a power-law of the form  $\mathrm{EM}(T)\propto T^{\alpha}$, with $\alpha$ the so-called emission measure slope \citep{jordan76,cargill94,cargill04}. In the nanoflare heating scenario,  the value of $\alpha$ is an indicator about how often a single strand is reheated and, hence, its observational inference helps to diagnose the heating frequency in nanoflare models. The method has been used to interpret active region core observations in terms of both high- and low-frequency heating \citep[see Table 3 in][and references therein]{bradshaw2012}. Overall, the values of the emission measure slope are in the range 2 to 5. The shallower slopes support low-frequency heating and the steeper ones are associated with high-frequency (or steady) heating. Low-frequency heating appears to be consistent with observed DEM distributions \citep{bradshaw2012,cargill14}. Quasi-steady nanoflare trains are consistent with a large percentage of observed active region cores \citep{reep13,cargill15}.


The analysis of the emission at different spectral lines with data from space instrumentation, such as the EUV Imaging Spectrometer (EIS) and X-Ray Telescope (XRT) onboard Hinode and the Atmospheric Imaging Assembly (AIA) onboard the Solar Dynamics Observatory (SDO), has demonstrated that well-constrained temperature measurements can be made over different coronal regions with varying physical conditions. \cite{warren12} performed a systematic study of the differential emission distribution in 15 active region cores. The results indicate that the temperature distribution in an active region core is often sharply peaked near 4 MK. When the properties of the emission measure distributions were compared to magnetic properties, such as the total unsigned magnetic flux, the results indicated that high-temperature active region emission is close to equilibrium, while weaker active regions seem to be dominated by 1 MK plasma evolving in the core.

A number of other investigations have analysed the diffuse emission from the cores of active regions by associating the slope to the heating frequency.  The overall results are rather inconclusive, with some active region cores showing a predominance of shallow slopes (consistent with low-frequency heating) and some others displaying steeper emission measure slopes (more consistent with high-frequency heating events). 
The most recent developments aim at applying machine-learning techniques to assess whether or not individual loops or strands are reheated before completely cooling \citep[see e.g.,][for recent applications]{barnes19,barnes21}.

\section{Coronal dynamics}

The solar corona is highly dynamic at a variety of time and spatial scales. The continuous emergence, cancellation, and dynamics of photospheric magnetic fields recycle the coronal field in time-scales of just a few hours \citep{close04}. Plasma flows, magnetodydrodynamic waves, and magnetic reconnection events are among the fundamental physical processes with relevance to the heating of the corona.  Magnetic reconnection is discussed in detail in Chapter 9. It is the process by which magnetic field lines break and reconnect, leading to an energy release that can heat and accelerate charged particles. We focus here on coronal plasma flows and magnetohydrodynamic wave activity.

\subsection{Coronal plasma flows}

Mass flows are among the three primary energy loss mechanisms in the solar atmosphere, together with radiation and thermal conduction \citep{withbroe77}. They are present in both magnetically open and closed coronal components. Open magnetic fields in coronal hole regions enable the continuous stream of particles in the form of fast solar wind, with speeds of 700\,--\,800 km~s$^{-1}$. The slow solar wind, with speeds of 300\,--\,500 km s$^{-1}$, is believed to originate from streamer belts and the edges of active regions \citep{mccomas08}. Observational signatures of large-scale plasma flows can be obtained from the analysis of white-light data from large-angle spectrometric coronagraphs, such as LASCO onboard SoHO. They allow to study the contribution of different structures to the brightness of the K corona \citep{wang07}. At smaller spatial scales, more detailed analyses are possible. They are based on the detection of Doppler velocity displacements, often correlated with measurements of the non-thermal broadening of coronal emission lines. From them, it has been found that persistent flows arise from the edges of active regions, which consist of a steady and quasi-stationary plasma stream \citep[see e.g.,][]{kojima99,sakao07,harra08,hara08}. Recent observations make use of multi-wavelength, high-resolution instrumentation to track plasma flows across different layers of the solar atmosphere and arising from regions with different magnetic activity level \citep[see][for example]{barczynski21,schwanitz21}.

Besides their relevance in connection with the solar wind, observations of plasma flows offer an important source of information on the physical characteristics, energy balance, evolution, and heating of active regions structures. The coronal plasma confined in magnetic loops is far from being in a static state.  Transient heating and cooling processes continuously fill and drain the material along the magnetic field. For example, asymmetric plasma heating in the foot-points of coronal loops creates siphon-type flows that are persistently detected in the form of red/blue-shifts in transition region UV spectral lines and as intensity variations in coronal emission lines. For instance, \cite{winebarger01} showed that brightness variations in loops that appear as static structures in imaging EUV observations with TRACE correspond to outflows with velocities between 5 to 20 km s$^{-1}$, interpreted as mass flows from the chromosphere into the corona. 

Spectroscopic measurements first obtained with SoHO/SUMER and then improved with Hinode/EIS have been of central importance in detailed studies of persistent loop plasma motions \cite[see Table 8.3 in][for a list of examples]{aschwanden19}. Measurements of plasma flow speeds in active regions range in between 5 and 100 km s$^{-1}$ \citep{harra08,delzanna08,doschek08,brooks2012b}. The measured Doppler flows in active regions show a mixed pattern with redshifts and blueshifts being stronger in cooler and hotter lines, respectively \citep{delzanna08}. 
Even when a particular region seems to be dominated by downflows, spatially localised outflows can be found \citep{doschek08}. The interpretation given to this is that the outflows might be directed along the long closed loops and/or the open magnetic fields. Obtaining Doppler velocity signatures of such flows is challenging because of a number of reasons. First, projection and integration along the line-of-sight affect their quantification. Also, the interpretation of the measured variations in terms of heating/cooling and/or flows is not straightforward. Finally, signatures of flows and waves are often diffult to distinguish from observations \citep{degroof04, verwichte10, demoortel15}.

Intensity variations are also seen to occur from the top of the loops towards the foot-points. First reported with the use of SoHO/EIT observations by \cite{degroof04}, they  were interpreted as coronal rain by \cite{muller05}.  Nowadays, the occurrence of flowing/falling plasma blobs is widely observed in the corona \citep{antolin10}. The phenomenon is believed to be due to the catastrophic cooling of the plasma and is linked to the mechanism of thermal instability. This in turn seems to be a consequence of foot-point-concentrated heating \citep[e.g.][, and references therein]{pelouze2022}. Coronal rain clumps fall with typical speeds in the range 30\,--\,150 km s$^{-1}$, that are considerably smaller than the free-fall velocity due to gravity. According to some numerical models, this is caused by the dynamical rearrangement of the coronal pressure, due to the increase in the pressure gradient that opposes gravity as the cool and dense plasma blob falls \citep{oliver14}.

In the last years, it has been realised that observed phenomena such as coronal rain and, in general, the cool plasma component of the corona, may be of significance to our understanding of the heating of the corona. Observations with instruments  onboard the Interface Region Imaging Spectrograph (IRIS) and the Solar Dynamic Observatory (SDO) imaging data indicate that long-period EUV intensity pulsations and periodic coronal rain are widely identified on large spatial scales of active regions and over long temporal scales with periods of minutes \citep{auchere14,sahin23}.  They are believed to originate from multiple evaporation-condensation cycles, produced by the joint action of thermal non-equilibrium and thermal instability. Their presence is believed to pose major observational constraints for coronal heating theories \citep{antolin22}.

\subsection{Coronal waves}

Early observations of solar coronal structures,  using coronal emission lines already pointed out the presence of brightenings in these structures. These brightenings were associated with propagating waves. For example, \cite{vernazza75} used coronal EUV lines to search for vertical propagation, such as revealed by time lags between brightenings at different heights, and found evidence for upward propagation of pulses. \cite{antonucci84} studied oscillations in the C II, O IV and Mg X UV emission lines observed with Skylab during a loop brightening and found periodic intensity fluctuations with a period of 141 s. These observations of coronal oscillations were restricted to time series analysis without any spatial information. They were based on the measurement of the temporal and spatial variation of spectroscopic properties (such as intensity, line width, and Doppler velocity) of coronal emission lines.  The situation changed drastically when high-resolution imaging and spectroscopic observations from instruments onboard e.g., SoHO, TRACE, Hinode, SDO, or Solar Orbiter spacecraft became available. Due to the temperature discrimination and spatial resolution of the EUV and soft X-ray telescopes onboard these satellites, observations of the solar corona have clearly demonstrated the existence of oscillations in the form of standing and propagating waves in solar coronal structures \citep{nakariakov05,demoortel05,aschwanden06,demoortel12}. In parallel, the development of increasingly sophisticated theoretical and numerical models has enabled to associate them to slow, fast, and Alfv\'en magnetohydrodynamic (MHD) waves \citep{roberts00,demoortel12}.  The observed wave dynamics may have some relevance in coronal heating processes \citep{arregui15a,vandoorsselaere20} as we discuss in the following.

\begin{figure}
 \title{Waves in extended regions of the corona}
 \includegraphics[width =6cm,angle=-90]{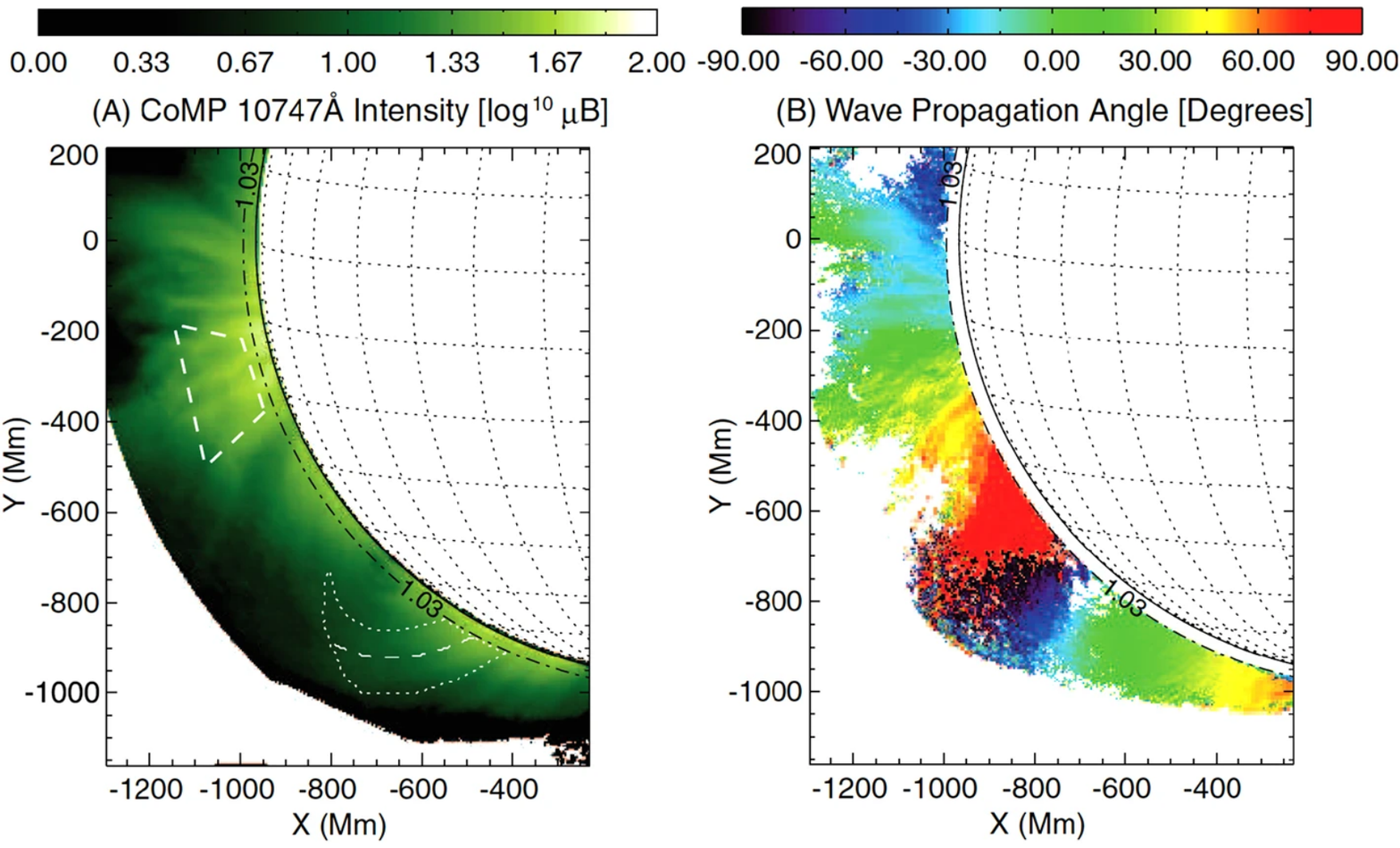}
\caption{Observational signatures of wave propagation in the form of images of the Fe XIII 10747 \AA\ line intensity (left) and Doppler shift in part of the corona observed by CoMP.  Adapted from \cite{tomczyk09}. From \cite{vandoorsselaere20}.}
 \label{fig:compwaves}%
 \end{figure}
 
A defining property of the wave dynamics is their ubiquity. Observations with the Solar Dynamics Observatory (SDO) have demonstrated that Alfv\'en waves are common in the transition region and corona and that they all seem to share a common origin \citep{mcintosh11}. Estimates about the energy carried by these disturbances vary, depending on the region of interest. The disturbances seem to be energetic enough to power the quiet corona and coronal hole regions, but not the active region corona. Observations with the highest available resolution instruments, confirm that wave activity  is all-pervasive in the EUV corona. \cite{morton13a}, in a study that combined Hi-C and SDO/AIA data, concluded that this activity was of low energy. \cite{lim2023} have recently extended the analysis to high-frequency waves (see Section~\ref{sect:acmechanisms}).
 
Decaying transverse coronal loop oscillations offered the first imaging evidence for standing waves in the corona associated with the periodic displacement of these structures \citep{aschwanden99, nakariakov99}. These large amplitude oscillations have periods of a few minutes and decay rapidly in time. They are caused by energetic events, such as nearby flares or coronal eruptions. Wave damping can be directly observed and measured. Information on periods and damping times has led to numerous developments in coronal seismology \citep{goossens02a,goossens06,arregui07a} and the development of models to explain the damping and possible wave heating (see Sections~\ref{sect:acmechanisms} and \ref{sec:acnummodels}). They are however rather sporadic events \citep{terradas18} and coronal loops are hot irrespective of the presence or absence of lateral displacements. 

Transverse coronal loop oscillations do not always decay in time. Sometimes, they are seen to persist for long periods \citep{tian12}, or even have their amplitudes grow in time \citep{wang12}. They consist of low-amplitude displacements measured with time-distance analyses of imaging EUV observations and some events display the coexistence of decaying and decayless oscillations in the same area \citep{nistico13}. The ubiquity of decayless oscillations \citep{anfinogentov13, anfinogentov15}, makes them a relevant component to be considered in the energy balance in the corona. A yet unidentified mechanism must be responsible to counteract the damping of the oscillations by supplying the required energy to the loops.

Observations made with the Coronal Multichannel Polarimeter (CoMP: \citealt{tomczyk08}) have demonstrated that coronal disturbances are present in extended regions of the corona at all times \citep{tomczyk07,tomczyk09}. Figure~\ref{fig:compwaves} displays their main features in the form of observations in intensity and Doppler shift measurements. The observed Doppler velocity fluctuations do not produce significant intensity variations and are interpreted as MHD kink waves propagating along the coronal magnetic field. They show signatures of in situ wave damping in the form of a discrepancy in the outward to inward wave power, but energy estimates seem to fall below the required amount to heat the ambient plasma. 
From the observational side, estimates exist for detected wave energy fluxes \citep[see Table 1 in][for an overview]{vandoorsselaere20}, yet the main difficulty lies in quantifying the fraction of that energy that gets dissipated. Still, the presence of wave dynamics in extended regions of the corona and their analysis is relevant in the context of both coronal seismology and coronal heating \citep{verth10,morton15,morton16,morton19,montesolis20,morton21}. Physical processes that may convert wave energy into heating are discussed in Section~\ref{sect:acmechanisms}.


\section{Coronal heating models}
For typical coronal densities of $10^9$~cm$^{-3}$ and a temperature of $10^6$~K, the radiative losses of the corona are estimated by $n^2P(T)$, where $P(T)\sim 10^{22}$erg~s$^{-1}$~cm$^{3}$ (see \citeauthor{rosner78}, \citeyear{rosner78} for a classical citation or \citeauthor{hermans2021}, \citeyear{hermans2021} for a more recent comparison of modern energy loss functions). Converting to SI units, we then have an estimated radiative loss of $L=10^{-5}$~W~m$^{-3}$. This coronal temperature and density lead to an estimate of the internal energy of the plasma as $\mbox{IE}=2\ 10^{-2}$~J~m$^{-3}$. From these two estimates, we could deduce an approximate cooling time $\tau_\mathrm{cool}$ of around 
\[ \tau_\mathrm{cool}=\frac{IE}{L} \approx 2000\ \mathrm{s}.\]
Thus, the corona would normally cool down in a matter of half an hour. This firmly points to the existence of a coronal heating mechanism, which keeps it at the observed temperature. Since the corona exhibits the high temperature in all regions, it implies that there must be a coronal heating mechanism operating in all coronal regions and at all times. These heating requirements are listed in Table~\ref{table:losses}. As stated before, the source of the coronal heating is very clearly in the photospheric convective motions. 

Coronal heating models are mostly classified in two groups, namely the DC mechanisms and AC mechanisms. The distinction between these groups of mechanisms is in the time scale of the aforementioned photospheric driver and its comparison to the Alfv\'en transit time. The Alfv\'en transit time is the time it takes to cross the particular coronal structure (loop, fibril, plume) with the fastest wave. The fastest wave in coronal structures is usually the Alfv\'en wave (aside from high plasma-$\beta$ regions, such as prominences or flaring regions, where gas pressure dominates magnetic forces and thermal conduction plays the major role). For a structure of length $L$ with an Alfv\'en speed of $V_\mathrm{A}$, the Alfv\'en transit time $\tau_\mathrm{A}$ is given by
\[ \tau_\mathrm{A}=\frac{L}{V_\mathrm{A}}.\]
Here the Alfv\'en speed is $V_\mathrm{A}=B/\sqrt{\mu\rho}$ with magnetic field strength $B$, density $\rho$ and magnetic permeability $\mu$. \par

If the time scale of the driver $\tau$ is slower than the Alfv\'en transit time $\tau_\mathrm{A}$: \[ \tau \gg \tau_\mathrm{A},\] it is a quasi-steady driving of the magnetic structure, which is called a DC heating mechanism. When the driver time scale is shorter than the Alfv\'en transit time \[\tau \ll \tau_\mathrm{A},\] mostly waves are launched in the structure, and these are then called AC heating mechanisms. 

\subsection{DC mechanisms}
\begin{figure}
\title{An illustration of Parker's braiding mechanism}
    \centerline{
        \includegraphics[width=\linewidth]{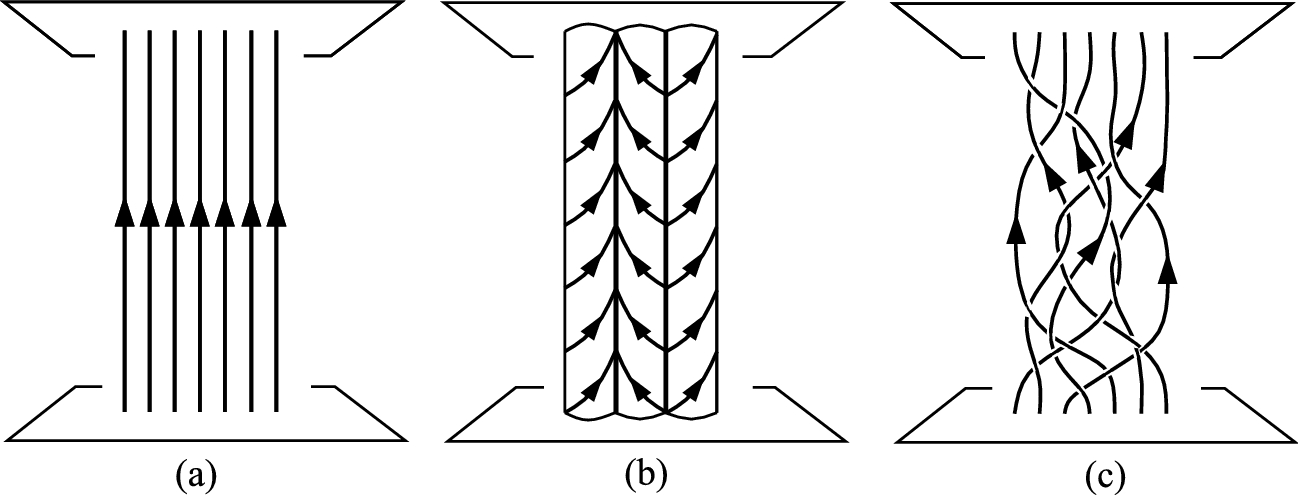}
    }
    \caption{The footpoint driving by the photospheric motions lead to a non-potential state of the magnetic field with currents (middle panel), and then leads to complicated magnetic structure (right panel). Figure taken from \citet{pontinLRSP}.}
    \label{fig:braiding}
\end{figure}

The lowest energy state of the coronal magnetic field is in its potential form. It is thus assumed that the magnetic field would return to its potential form if it were not driven from the photospheric convection. The photospheric convection then leads to the creation of a non-potential form of the magnetic field. This has two consequences: (1) the non-potential fields have a slow buildup of currents, and (2) the driving leads to a tangling or braiding of the magnetic field. The latter effect is also known as Parker braiding \citep{parker1972,pontinLRSP}, and is displayed in Fig.~\ref{fig:braiding}.
With the buildup of currents, the non-potential magnetic energy starts being dissipated by ohmic friction. However, it is thought that this {\em ohmic dissipation} of the currents is insufficient to heat the solar corona, because the coronal magnetic Reynolds number $R_\mathrm{m}$ is very high \citep[$R_\mathrm{m}\sim 10^{12}$, see][]{terradas18}. The ohmic dissipation happens on a dissipative time scale $\tau_\eta$, given by \citep{roberts1967}
\begin{equation}
    \tau_\eta=R_\mathrm{m} \tau_\mathrm{A}.
\end{equation}
This shows that the ohmic dissipation of the built-up current is extremely slow for typical Reynolds numbers. 

Because of the slow ohmic dissipation, a lot of non-potential energy can be stored in the coronal magnetic field. The magnetic field configuration becomes increasingly complex. It is thought that that strongly tangled magnetic field eventually reconnects in what is called a {\em nanoflare}. These are (as yet unobserved and unobservable) phenomena which are scaled-down versions of flares. The nanoflares are caused by a sudden decrease of the Reynolds number: either due to strong converging flows creating a stronger forcing and small lengths scales, or anomalous resistivity attributed to instabilities occurring in the current sheet which effectively lower the Reynolds number several orders of magnitude. 

Since the nanoflares are highly localised phenomena and the thermal conduction is practically only along the magnetic field, the heating by the nanoflares is then spread on single field lines. Thus, it is thought that the scenario of nanoflare heating of the corona leads to the existence of multi-stranded loops, where each individual field line has its own independent thermodynamic evolution \citep[e.g.][]{aschwanden2000,viall2013}. On the other hand, nowadays it is thought that the strand's evolution cannot be entirely independent \citep{magyar2016}, given that they are continuously mixed by the Kelvin-Helmholtz instability operating in transversely oscillation loops \citep{terradas2008b}. Still, despite this criticism, multi-stranded loops models are a popular topic in the coronal heating community (for more detail, see Section~\ref{sec:models}), and the effect of multi-strandedness is both considered in DC heating models \citep[e.g.][]{reid2023} and AC heating models \citep{vd2008b,luna2010,guo2019}.

An important concept in nanoflare heating of the solar corona is the distribution of nanoflares as a function of energy. The reason is that it is postulated that the nanoflares are occurring as scaled versions of regular flares, with a continuous scaling up to some minimum energy. Following \citet{hudson1991}, we can write the histogram $N(E)$ of number of flares as function of energy $E$ in a power law form:
\begin{equation}
    N(E) \sim E^{-\alpha}. \label{eq:dcpowerlaw}
\end{equation}
The total heating $H$ due to nanoflares can then be calculated by the integral 
\begin{equation}
    H = \int_{E_\mathrm{min}}^{E_\mathrm{max}} N(E) E dE,
\end{equation}
where $E_\mathrm{max}$ is the maximum flare energy and $E_\mathrm{min}$ is the minimum flare energy. The value for the latter (and for the former, for that matter) is currently not know, neither theoretically nor observationally. \\
With the previous power law form, the integral can be evaluated to be 
\begin{equation}
    H_\mathrm{DC} \sim \left.\frac{E^{2-\alpha}}{2-\alpha}\right\vert_{E_\mathrm{min}}^{E_\mathrm{max}}.
\end{equation}
In case $\alpha<2$, the heating is dominated by the high energy flares, but given their localised nature in active regions (which is in contrast to the corona which is also hot outside of active regions) they are not able to the heat the corona to its observed temperature. Thus, a value of $\alpha>2$ is desirable if you like to believe in the nanoflare heating of the corona. In the latter case, the heating is dominated by the small-scale nanoflares, which presumably occur everywhere and therefore heat the entire corona (in active regions and quiet Sun alike). 

\begin{figure}
\title{Nanoflare energy spectra}
    \includegraphics[width=.8\linewidth]{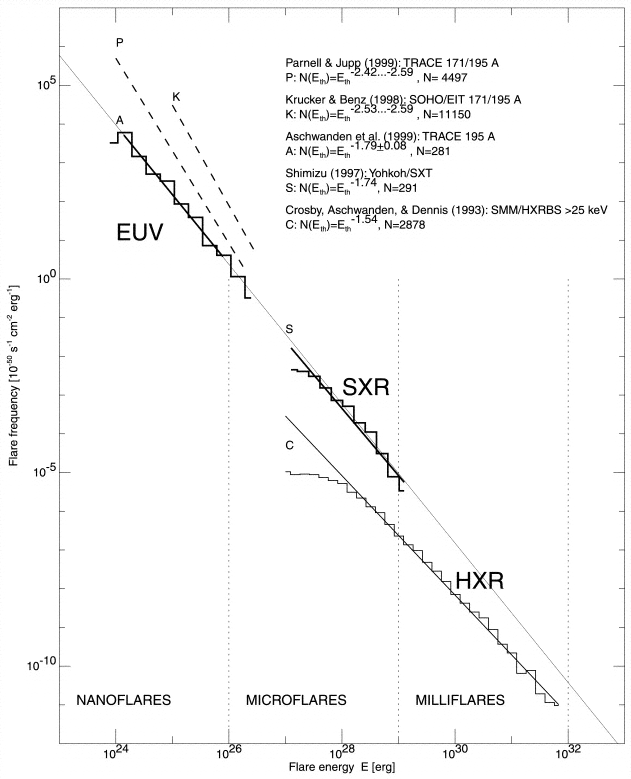}
    \caption{Observed nanoflare energy distribution from different instruments and the associated power law indices are indicated in the legend of the figure. Figure taken from \citet{aschwanden2000b}.}
    \label{fig:nanoflares}
\end{figure}

The measurement of the power law index of the nanoflare distribution $N(E)$ was a big topic around the turn of the century \citep{aschwanden2000b,parnell2000}. The distribution of nanoflares from different instruments is shown in Fig.~\ref{fig:nanoflares}. The currently obtained values for the power law index $\alpha$ range between 1.8\,--\,2.7 \citep[e.g.][and references therein]{purkhart2022} and are thus inconclusive if nanoflares play a major role in the heating of the corona.

\subsection{AC mechanisms}\label{sect:acmechanisms}
When the driving time scale $\tau$ is shorter than the Alfv\'en transit time $\tau_\mathrm{A}$, the shuffling of the magnetic field happens on fast time scales. Rather than creating tangling, this then results in the generation of (magnetic) waves. In an unstructured corona, these waves are Alfv\'en and fast waves, but with the structuring imposed by the density contrast in loops or plumes, additionally kink waves are also excited.\\ 
Kink waves are transverse motions of an overdense flux tube \citep{zaitsev1975,wentzel1979,edwin1983}. They are like surface Alfv\'en waves (transverse waves of an interface), but warped onto a cylindrical (or any other closed) surface. They are driven mostly by magnetic tension \citep{goossens2009}, have a large parallel vorticity \citep{goossens2019}, but also have compression and magnetic pressure variations. 

As for DC heating mechanisms, the waves may contain sufficient energy to heat the solar corona, but the problem is that it is dissipated on long, resistive time scales $\tau_\eta$. Thus, it is necessary to have mechanisms to transport the wave energy from the large scales to the small scales, where the Reynolds number is sufficiently low. Over the course of the years, several mechanisms have been proposed for that. 

\begin{figure}
\title{A simulation result of phase mixing}
    \includegraphics[width=.8\linewidth]{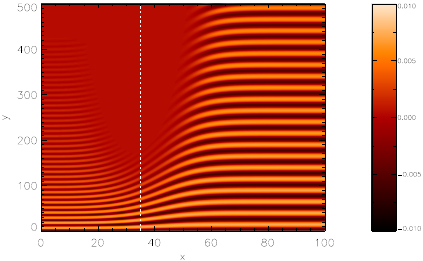}
    \caption{The colour scale shows Alfv\'en waves propagating in a non-uniform plasma, showcasing the effect of phase mixing. Figure taken from \citet{mclaughlin2011}.}
    \label{fig:phasemixing}
\end{figure}

The first such mechanism to consider is {\em phase mixing} \citep{heyvaerts83}. The mechanisms for phase mixing is usually referred to as the phase mixing of Alfv\'en waves, but the mechanism works equally well for other wave types. The crucial ingredient is the spatial variation of the phase speed of the wave. From a uniform or large-scale driver, the wave fronts on different field lines gradually get out of phase, leading to small length scales. This is shown in Figure~\ref{fig:phasemixing}. The wave fronts start from the bottom of the figure as a straight line, but bend due to the Alfv\'en speed gradient. Then the central part of the figure shows the decrease of wave power, because there the Reynolds number becomes sufficiently low to damp the Alfv\'en waves. One of the inherent problems of phase mixing is that the wave damping (and the resulting heating) does depend on the transport coefficients for viscosity or resitivity \citep[see][for analytical expressions]{roberts00}, which are extremely small in the solar corona. 

Another popular mechanism is called {\em resonant absorption} of kink waves \citep{chen1974}. Also for this wave damping mechanism, the presence of an Alfv\'en speed gradient is crucial. Typically, the mechanism is modelled in a cylindrical configuration mimicking a loop, with a smooth variation of the density profile across the loop's edge. In such a system, a transverse wave is found and is called a {\em kink wave}. It has a transverse displacement of the whole loop body, in a nearly incompressible way, which has a frequency 
\begin{equation}
    \omega_\mathrm{k}=\sqrt{\frac{2B^2}{\mu(\rho_\mathrm{i}+\rho_\mathrm{e})}}.
\end{equation}
Here $\rho_\mathrm{i}$ and $\rho_\mathrm{e}$ are the densities inside and outside the loop respectively \citep[for a derivation, see e.g.][]{edwin1983}. That kink frequency $\omega_\mathrm{k}$ is an average of the interior and exterior Alfv\'en frequency. Thus, at some resonant point, the kink frequency must equal the local Alfv\'en frequency. Thus, at that resonant point, the loop's global motion is converted from the large scale to small scale, shearing Alfv\'en waves similar to phase mixing \citep{soler2015}, allowing to let resistive heating pick up again. This mechanism has the advantage that the damping of the global kink mode happens on a short time scale that does not depend on the resistivity. However, the wave energy in the Alfv\'en waves at the resonant point only damps at resistive time scales \citep{terradas18}.

A third possible mechanism is the creation of turbulence from non-linear wave interaction. The development of turbulence comes in essence from the (in)famous non-linear term\footnote{This term is also the subject of the millennium problem on the existence and smoothness of the solutions to the Navier-Stokes equations.} in the fluid equations $\vec{v}\cdot\nabla\vec{v}$. Imagine driving the system with a single frequency $\omega$. The non-linear term is then acting as a sink at the frequency $\omega$, but acting as a source at the Fourier component with the double frequency. From that double frequency, one can progressively fill a whole Fourier spectrum with wave power. 

\begin{figure}
\title{Structure of solar wind wave-power spectral range}
    \includegraphics[width=.9\linewidth]{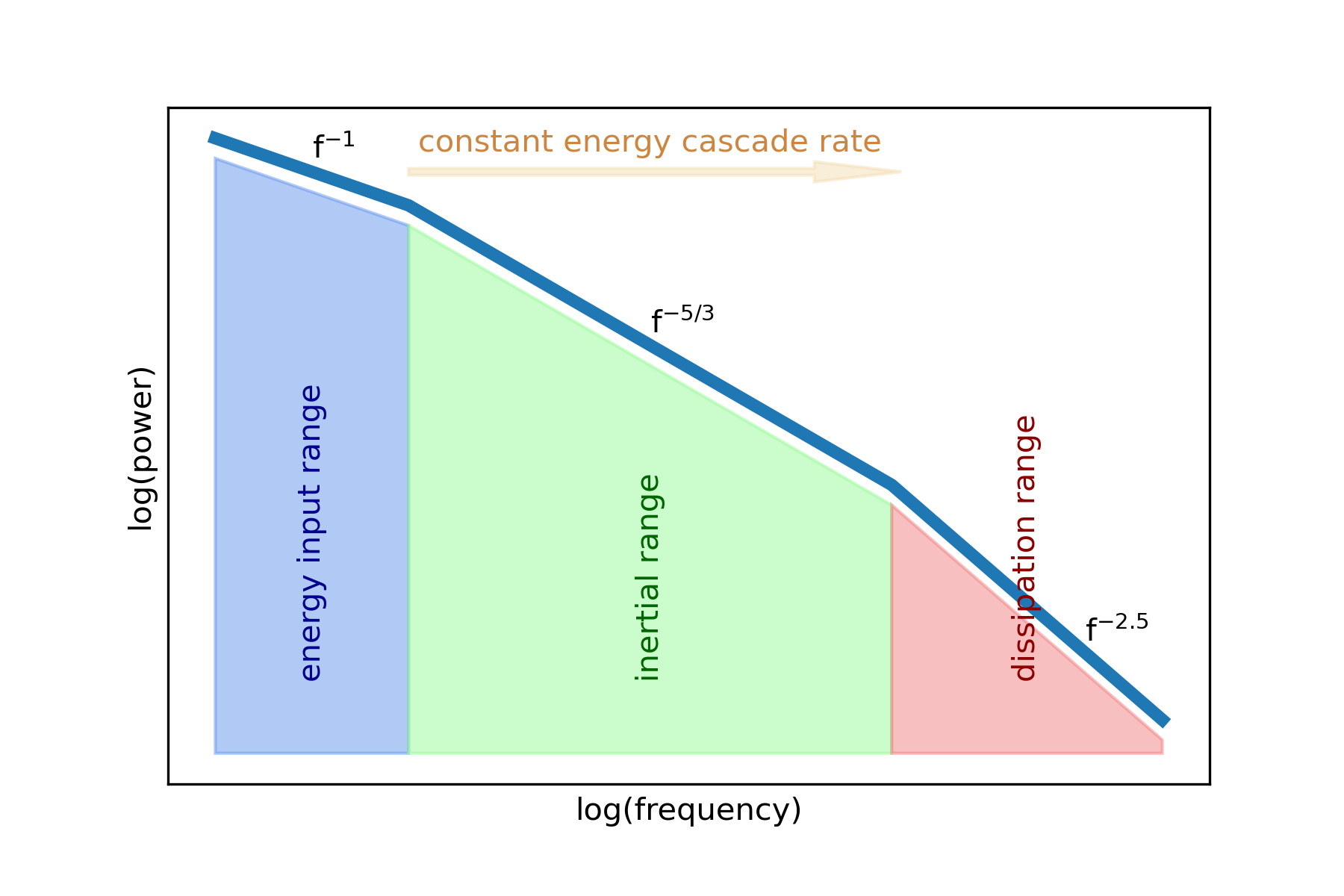}
    \caption{The formation of a power law due to turbulence, with the energy injection scale, the inertial range and the dissipation scale. Figure taken from \citet{vandoorsselaere20}.}
    \label{fig:powerlaw}
\end{figure}
In the spectral domain, the wave power arranges itself into a power law behaviour. This is readily observed in the solar wind \citep{bruno2013}, and it is likely that such a power law also exists in the solar corona \citep[e.g.][]{tomczyk07,morton15}. In the solar wind, the spectral range is broken into 3 (or more) regimes, see Figure~\ref{fig:powerlaw}. The large-scale regime is called the energy injection scale, where presumably coronal waves are excited at granular or supergranular scales. This regime has a power law index of -1. Then there is the inertial range, where the power law index is -5/3. This can be derived in fluid dynamics from the assumption that the energy cascade rate in this inertial range is independent from the scale, as was first pointed out by \citet{kolmogorov1962}. When the turbulence reaches a small scale, the Reynolds number becomes low and dissipation sets in to damp the fluctuations and heat the corona. Depending on the dissipation mechanism, the power law index there is typically between -2 and -3. 

In the corona, we believe the turbulence develops from a number of mechanisms. The first one is the interaction of counterpropagating Alfv\'en waves, as it is also believed to be responsible in the solar wind. During the interaction of the Alfv\'en wave, they have a short time to non-linearly modify the other wave front, leading to turbulence. A caveat in coronal structures is that the Alfv\'en waves need to have a variation along their direction of polarisation \citep{shestov2022}, which necessarily lead to non-zero azimuthal wave numbers $m$ in magnetic cylinders.\\
A second method to create turbulence is through the Kelvin-Helmholtz instability at the shearing boundary of loops oscillating with the kink wave \citep{terradas2008b}. For such standing kink waves, the magnetic shear (i.e. current) is near the footpoints, and does not stabilise the velocity shear. Thus, the classical Kelvin-Helmholtz instability (KHI) takes place because of the rapid change of azimuthal velocity at the loop's boundary, evolving in what is sometimes called Transverse Wave Induced Kelvin-Helmholtz (TWIKH) rolls. The velocity shear is enhanced by resonant absorption \citep{antolin2019} taking place at the same location. \\
The third mechanism for generating turbulence in the solar corona is called uniturbulence \citep{magyar2017}, operating in propagating kink waves. In that mechanism, turbulence is not generated from the KHI, because there is a stabilising magnetic field that inhibits the KHI growth. Instead, the kink wave is self-interacting \citep{magyar2019} leading to the formation of small scales, kink wave damping and heating \citep{vd2020}.

Nowadays, researchers are starting to consider the AC mechanisms also through similar power law descriptions as the DC heating mechanisms (see Equation~\ref{eq:dcpowerlaw}). As for the DC heating mechanisms, we may say that the power of AC heating events $P(\omega)$ follows a power law as a function of the frequency $\omega$: 
\begin{equation} P(\omega) \sim \omega^{-\delta}.\end{equation}
As for the DC heating mechanisms, we also have to integrate this over the whole frequency domain to find the entire AC heating energy input $H_{\rm AC}$ into the corona:
\begin{equation} H_\mathrm{AC} \sim \int P(\omega) d\omega = \left. \frac{\omega^{1-\delta}}{1-\delta}\right|^{\omega_{max}}_{\omega_{min}}. \end{equation}
In contrast to the DC heating mechanism, the critical value for the slope is 1. Such considerations are starting to be considered for modern solar telescopes, in casu SolO/EUI, SDO/AIA, (u)CoMP \citep{morton19,lim2023}. In the latter work, it is tentatively concluded that for small structures, $\delta<0$ so that the high frequency waves have the most important contribution in the AC heating.

\section{Numerical models of coronal heating}\label{sec:models}
\subsection{Numerical models of DC heating}
In numerical models of DC heating, there are basically two groups. The first group consider a coronal loop as the hydrodynamic evolution of 1D strands after nanoflare heating. The second group tries to capture the full corona at a large scale, following an ab-initio approach. 

In the first group of models, the main point that is considered is that nanoflares are very localised heating events. They would suddenly dump their energy on a specific field line, after which the plasma on that field line would evolve hydrodynamically, with the field only providing the geometry but do not restrict the physics. \\ 
An example of results from such 1D codes are found in \citet{bradshaw16}. In that work the loop (bundle) is modelled as 400 independently evolving field lines. On these field lines, a heating function is chosen and parametrised (e.g. continuous heating mimicking waves, or impulsive, stochastic nanoflares). Despite the seemingly simple nature of the model, the thermodynamic evolution is often non-intuitive. The superposition of different strands allows to explain several observational phenomena: for instance in the aforementioned paper, it was shown that these 1D models are compatible with the observed cooling behaviour of coronal loops \citep{viall12}. \\
Because of their success, such 1D models for coronal loops are also popular in explaining the observed behaviour of flares. Still, critical voices exists saying that the KHI of loops oscillating with kink waves will rapidly mix such individual strands, effectively creating cross-field coupling in the loop \citep{magyar2016}.

The second group of models consist of 3D numerical models of a part of the solar corona. They all  basically follow the lead of \citet{gudiksen2005}. Such simulations typically take a simulation of the solar convection zone as the bottom boundary condition. Then they let the plasma in the corona evolve self-consistently, driven by the photospheric simulation. It is found that the corona self-organises into loops, whereby it is currently debated what a coronal loop actually is \citep{bingert2011,malanushenko2022}. These works seem to point out that the loop is determined as the line-of-sight integration of plasma structures, rather than a magnetic flux tube. \\
The heating in such 3D models is naturally due to the lower Reynolds numbers. In most simulations, the heating is due to numerical resistivity. In the more modern iterations \citep[e.g. using MURAM,][]{rempel2017}, a strong weight is given to the compressive viscosity. However, the differentiation between the numerical resistivity or viscosity does not seem to have a large effect on the apparent structure of the resulting corona. \\
Such models are starting to be used to characterise the nature of the nanoflares. For instance, \citet{kanella2019} quantify the distribution of the Joule heating events in such a simulation, to understand the spatial distribution and occurrence rates of nanoflares. This is an avenue that could be pursued in the future, providing constraints on the possible coronal heating mechanisms in such 3D simulations. 

\subsection{Numerical models of AC heating}\label{sec:acnummodels}

Early numerical models considered the heating of coronal loops by resonant absorption in compressible, resistive MHD. The loops were approximated as cylindrical plasma columns and were excited by an external periodic driver \citep[see e.g.,][]{poedts89a, poedts89b}. The driven system reaches a stationary state. Part of the energy from the incident waves is extracted by the coronal loops, gets dissipated, and can heat the plasma.  Resonant absorption is very efficient for typical coronal loop parameters as a considerable part of the energy supplied by the external source is dissipated ohmically and converted into heat  \citep{poedts90b, poedts90a}.  The process works for wave modes with any azimuthal wavenumber $m$, not only kink modes ($m=1$) and can thus play a role in the absorption of p-modes with high azimuthal wave numbers by sunspots \citep{goossens92}. As shown by \cite{poedts90c}, resonant absorption evolves in time with phase mixing towards the resonant point. 

For phase mixing, the numerical models were previously mostly in 2D \citep{mclaughlin2011}. These heating models were in highly idealised setups, but recently such models started to include the lower atmosphere \citep{vandamme2020}. Moreover, the transition was made to 3D. In 3D the effects were considered of driving with data-inspired power laws \citep{pagano2019}. Moreover, the latest simulations significantly go beyond simplified magnetic field configurations \citep{howson2020}. The latter considered wave propagation and damping in a 3D configuration induced by magnetic braiding, which offered readily available small scales for enhanced wave dissipation. The conclusion of most of these simulations on phase mixing heating remains the same: it is hard to heat the corona, unless high wave driver amplitudes are considered, in addition to large resistivity (low Reynolds numbers). Thus, it seems that even the small scales introduced by the magnetic braiding do not facilitate heating by phase mixing enough to heat the corona. 

Another group of models considers the heating by torsional Alfv\'en waves. These find their root as early as \citet{moriyasu2004,buchlin2007,antolin2008}. They have 1D field line models for solar coronal loops on which torsional Alfv\'en waves are driven at the loop's footpoint. The heating is then provided by either the steepening of the Alfv\'en waves and associated dissipation, or Alfv\'en wave interaction forming turbulence. A famous example of the latter is \citet{vanballegooijen2011}, who considers the turbulent interaction of a superposition of Alfv\'en waves in close loop models. Given enough wave driving at the footpoint, the wave heating can create loops with a coronal temperature, even leading to condensations as a consequence of very strong heating. The latest incarnation of this model is \citet{matsumoto2018}, who extended this model to a 3D system. The main issue with these models is that they invariable ignore radial structuring of coronal loops, either by using reduced MHD, or by considering a uniform cross-section before the launching of the waves. Thus, it is rather unclear how such loop models correspond to observations, which are actually density-enhanced with respect to the background coronal plasma. 


In the last couple of years, there has been a leap into models which do the heating with the KHI and the associated TWIKH rolls \citep{terradas2006b,antolin2013}. It has been found that coronal loops get fully turbulent \citep{karampelas2018,antolin2019}. That leads to significant heating, even as much as being able to compensate the loop's radiative losses \citep{shi2021}. This avenue seems promising, but for now, no observational confirmation exists \citep[aside from the circumstantial][]{antolin2015,pascoe2020}. Still, the predicted damping \citep{vd2021} seems to be compatible with the observed dependence of the damping with the amplitude \citep{nechaeva2019,arregui21}. On the other hand, the loop heating is only successful for low density loops, but not for higher density loops or driving at a non-resonant frequency \citep{demoortel2022}. Additionally, it does not provide a convincing mechanism to provide mass to the loop from an initially homogeneous corona, and thus does not constitute a self-consistent mechanism for the formation of the loop. 

A final group of wave heating models is by uniturbulence \citep{magyar2017}. This wave heating mechanism is operating in coronal open regions \citep{magyar2021}, where the coronal plumes may provide the structure for creating the uniturbulence. The damping of the transverse waves is significant \citep{vandoorsselaere20,morton21}, and the resulting turbulent heating must contribute something to the energy budget, on top of the heating by Alfv\'en wave turbulence from reflected Alfv\'en waves. This route is interesting to pursue, because it seems from WKB models of the solar wind \citep{vanderholst2014} that there is insufficient wave driving in coronal holes. As for the KHI heating models, also the observational evidence for uniturbulence is uncertain, although \citet{pant2019} can self-consistently explain the observed relationship between Doppler shift and spectral line width \citep{mcintosh2012}. This also matches well with the observed non-thermal broadening in the corona and solar wind. 

\section{Summary and Outlook}
Imaging and spectroscopic observations display a multi-million degree solar corona that cannot be explained by means of thermal energy transport mechanisms.  EUV and X-ray emission, primarily produced by highly ionised atoms in the corona, is highly structured by the magnetic field and its overall appearance varies with the solar-activity-cycle. On the large-scale, the corona is the source of the solar wind. The high temperatures in the corona help to accelerate these particles to high speeds, which can now be measured by spacecraft in the solar wind. At the active-region scale, detailed analyses of intensity variations and thermal structure provide constraints to coronal heating theories and model-specific parameters, such as the frequency of small-scale heating events. The strongly dynamic nature of the corona makes plasma flows, wave dynamics, and magnetic reconnection relevant components in the mass and energy balance of the corona. 

Concerning coronal heating models, it is safe to say that wave heating models made a leap in the last 5 years from 1D models or even cartoon-like models to modern 3D configurations. Despite this step forward, it is still not straightforward to heat coronal loops to the desired temperature. For instance, \citet{shi2021} show that a coronal loop with density $10^8$cm$^{-3}$ can be heated to coronal temperatures and sustain it against radiative losses. However, \citet{demoortel2022} have shown that this only works for a specific set of parameters (low density, resonant driving). Thus, new ingredients are needed to make AC heating models work. For example, multi-mode driving \citep{guo2019b,afanasyev2020} and inherent fine structure \citep{guo2019,howson2020} will be needed to enhance the heating. 

The Extreme Ultraviolet Imager onboard ESA's Solar Orbiter (SolO/EUI) is already offering remarkable data on small-scale dynamical phenomena, such as the relaxation of braided coronal magnetic fields through magnetic reconnection \citep{chitta22}, quiet-Sun intermittent brightening events \citep{berghmans21}, or high-frequency spatially resolved wave dynamics \citep{petrova23, lim2023,zhong23}.\\
The Marshall Grazing Incidence X-ray spectrometer (MaGIXS) is a slitless imaging spectrograph that
enables to obtain spatially resolved soft X-ray (6\,--\,24 \AA) spectra across a wide field of view. The first science results, described in \cite{savage23}, indicate that the analysis of spectrally dispersed soft X-ray spectral imaging can help in determining the spatial and temporal properties of coronal heating events.\\
The NASA Multi-slit Solar Explorer (MUSE) mission, composed of a multi-slit EUV spectrograph and an EUV context imager in several narrow wavelength-bands, will provide spectral and imaging diagnostics of the solar corona at high spatial (<0.5 arcsec), and temporal resolution (down to ~0.5s). This is expected to shed light into the physical processes that drive coronal heating \citep{depontieu22}. Simultaneous remote and local observations of the out-flowing coronal plasma, obtained with both Solar Orbiter/Metis and the Parker Solar Probe instruments, are now possible, as shown by \cite{telloni23} who report observational estimates of the heating rate in the slowly expanding solar corona.

These are just a few examples of advances that are expected to help us refining our coronal heating models and improving our understanding of the solar corona.

\bibliography{ch}


\Backmatter

\bibliographystyle{elsarticle-harv}
 
\end{document}